\documentclass[conference]{IEEEtran}

\IEEEoverridecommandlockouts
\usepackage{cite}
\usepackage{amsmath,amssymb,amsfonts}
\usepackage{algorithmic}
\usepackage{graphicx}
\usepackage{textcomp}
\usepackage{xcolor}
\usepackage{caption}
\usepackage{tikz}

\def\BibTeX{{\rm B\kern-.05em{\sc i\kern-.025em b}\kern-.08em
    T\kern-.1667em\lower.7ex\hbox{E}\kern-.125emX}}
\begin{document}

\title{Towards Improving the Trustworthiness of Hardware based Malware Detector using Online Uncertainty Estimation\\
\vspace{-0.6cm}
{\footnotesize \color{red}To be presented at DAC 2021.}
\vspace{-0.6cm}
}

\author{\IEEEauthorblockN{Harshit Kumar, Nikhil Chawla, Saibal Mukhopadhyay}
\IEEEauthorblockA{\textit{Dept. of Electrical and Computer Engineering} \\
\textit{Georgia Institute of Technology}
Atlanta, USA \\
hkumar64@gatech.edu, nchawla6@gatech.edu,  saibal@ece.gatech.edu}

}

\maketitle


\begin{abstract}
Hardware-based Malware Detectors (HMDs) using Machine Learning (ML) models have shown promise in detecting malicious workloads. However, the conventional black-box based machine learning (ML) approach used in these HMDs fail to address the uncertain predictions, including those made on zero-day malware. The ML models used in HMDs are agnostic to the uncertainty that determines whether the model ``knows what it knows," severely undermining its trustworthiness. We propose an ensemble-based approach that quantifies uncertainty in predictions made by ML models of an HMD, when it encounters an unknown workload than the ones it was trained on. We test our approach on two different HMDs that have been proposed in the literature. We show that the proposed uncertainty estimator can detect $>90\%$ of unknown workloads for the Power-management based HMD, and conclude that the overlapping benign and malware classes undermine the trustworthiness of the Performance Counter-based HMD.  
\end{abstract}


\begin{IEEEkeywords}
Machine Learning, uncertainty, security
\end{IEEEkeywords}

\section{Introduction}


The application of ML models in Hardware-based Malware Detectors (HMDs) have recently gained interest \cite{demme_performance_counter, 8280556,8465828,8192483,chawla_iot}. HMDs detect malware by capturing the invariant functionality of malware at the hardware level. HMDs based on features derived from performance counter (HPC) data \cite{demme_performance_counter, 8280556,8465828}, electromagnetic emissions \cite{8192483}, and power-management signatures \cite{chawla_iot} have been demonstrated to classify workloads as benign and malware. Snapdragon processors from Qualcomm are using HMDs for online malware detection \cite{noauthor_qualcomm_2017}.

\begin{figure}[!b]
    \centering
    \includegraphics[width = 0.5\textwidth]{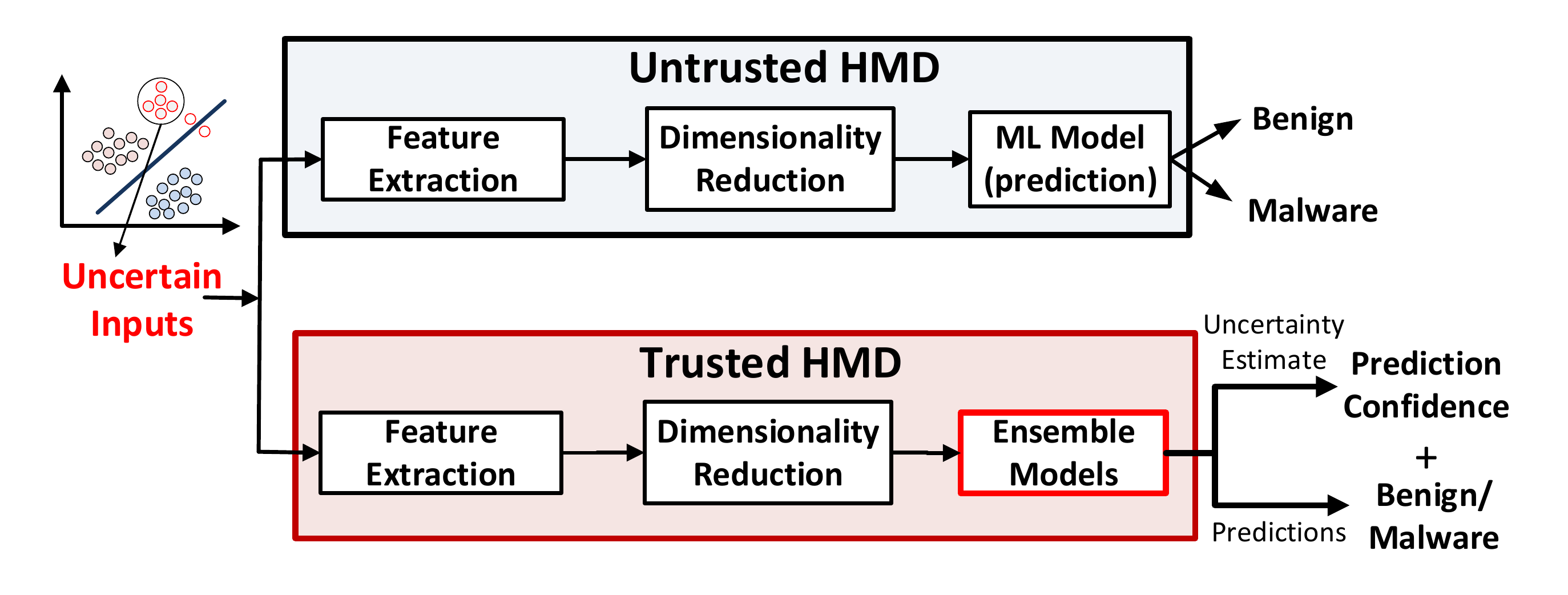}
    \caption{Untrusted HMD (Conventional approach) vs. Trusted HMD (Proposed approach)}
    \label{fig:fig1}
\end{figure}

While machine learning (ML) algorithms employed in HMDs work well in general, their black-box nature obfuscates their decision-making process. The HMDs proposed in the literature follow a general pipeline of the black-box approach focused on increasing the average predictive accuracy. An ML model is created for classifying a workload into benign or malware, and a learning algorithm is used to fit the model's parameters to optimize the accuracy on the available dataset which can be HPC, EM, or power management based signature. The result of the model prediction is always binary, \textit{i.e.}, benign or malware, shown as Untrusted HMD in Fig. \ref{fig:fig1}. 

If an ML-based HMD encounters a signature generated by a zero-day malware that wasn't part of its training data, it will make a prediction irrespective of the fact that the model knows nothing about the malware. Moreover, the choice of the hardware-sensors may create overlapping classes that affects the model's confidence on predicting data points in the region of overlap. If such erroneous predictions go unchecked, then it can severely undermine the efficacy of the HMD. Therefore, there is a need to measure the trustworthiness of the ML models in HMDs, as it facilitates treatment of uncertain inputs and special cases explicitly (shown as Trusted HMD in Fig. \ref{fig:fig1}). For example, in the case of zero-day malware, we can collect forensic data to re-train the model on this new class of malware, and pass the uncertain prediction to a security specialist for further inspection. 

Uncertainty of an ML model aims to quantify the limits of the model’s knowledge and provide us an insight into the confidence of the model while making predictions. Consequently, it plays a pivotal role in ensuring the robustness and trustworthiness of the ML models deployed in diverse applications \cite{kendall2017uncertainties,lakshminarayanan2017simple}. While the notion of uncertainty as a measure of trustworthiness is well known in the machine learning and deep learning community, these concepts have not been applied to the field of HMD. Prior works on uncertainty quantification in ML models are focused on deep learning and mostly uses Monte-Carlo sampling methods like dropout \cite{gal2016dropout}. However, such approaches are not directly applicable to traditional ML models like SVM, Random Forest, and Logistic Regression used in an HMD. This motivates us to come up with a ``general-purpose solution" using ensemble based framework, that can deliver uncertainty estimates with minor modifications to the standard training and testing pipeline. To the best of our knowledge, the HMD community has not addressed this important problem.

\noindent We make the following contributions in this paper: 
\begin{itemize}
        \item  We provide a background on the different kinds of uncertainties and their role in determining the efficacy and robustness of an ML model in a HMD.
    \item We evaluate the trustworthiness of HMDs by addressing domain-specific problems like zero-day malware through the use of predictive uncertainty. 
    \item For this purpose, we formulate an ensemble based framework to estimate uncertainty that can be easily incorporated in regular ML training and testing pipeline.

    \item We evaluate our framework on two different HMDs proposed in the literature, and present empirical results. Our results highlight the importance of including uncertainty analyses while evaluating possible HMD candidates. 
\end{itemize}

The rest of the paper is organized as follows. We provide a brief background in Section \ref{sec:background}. Next, we explain our proposed framework in Section \ref{sec:proposed_methodology}. We present our evaluation setup and datasets in Section \ref{sec:eval_method} followed by the results and discussion of our evaluation in Section \ref{sec:eval_results} and conclude in Section \ref{sec:conclusion}.

\begin{figure*}[t]
   \begin{minipage}{0.6\textwidth}
     \centering
     \includegraphics[width=\linewidth]{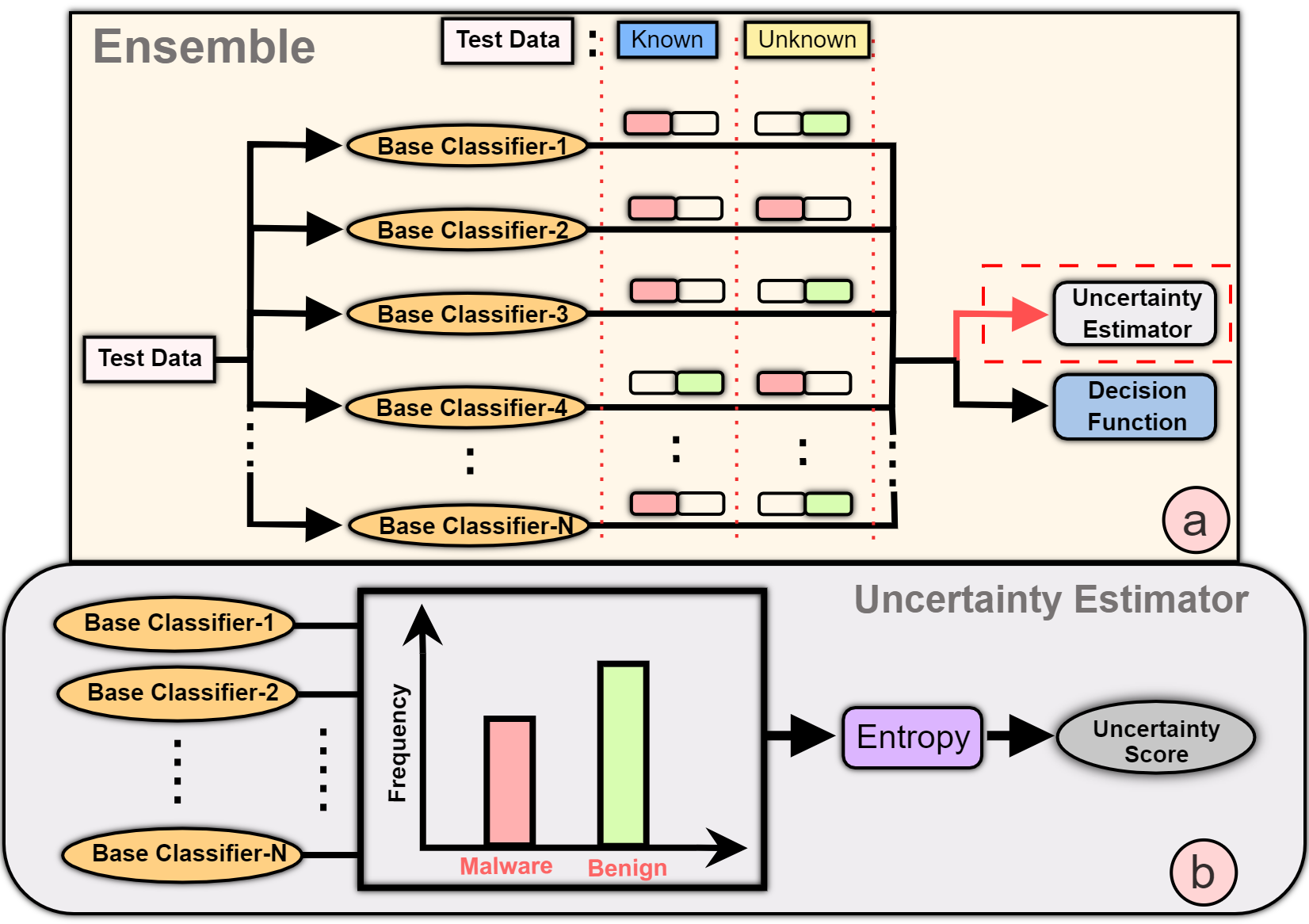}
     \caption{Overview of the proposed framework.}\label{Fig:overview}
   \end{minipage}\hfill
   \begin{minipage}{0.4\textwidth}
     \centering
     \includegraphics[width=0.82\linewidth]{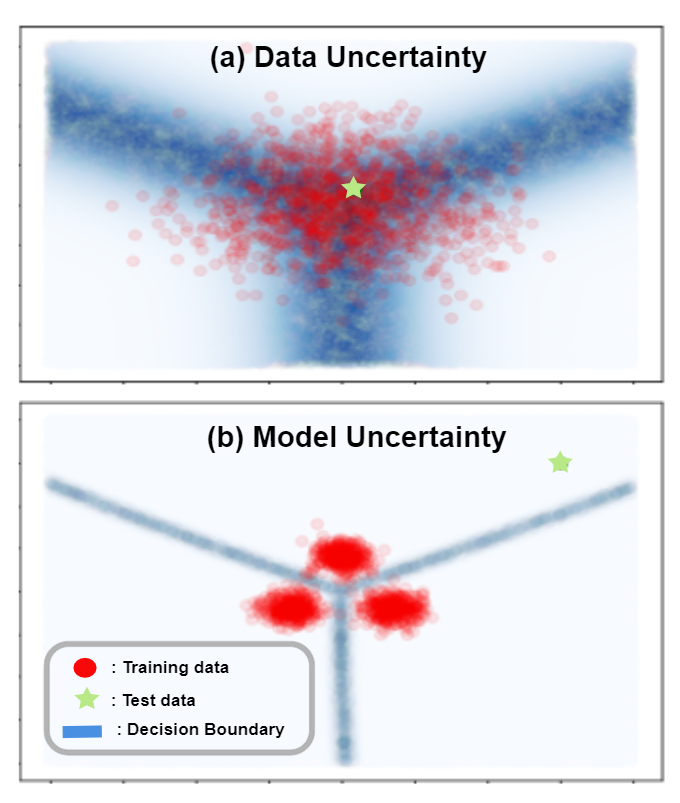}
     \caption{Types of uncertainty \cite{malinin2018predictive}}\label{Fig:uncertainty}
   \end{minipage}
\end{figure*}

\section{Background}
\label{sec:background}

\subsection{Uncertainty}
Bayesian modeling involves two kinds of uncertainties \cite{kendall2017uncertainties}. \textbf{Alleatoric uncertainty or Data Uncertainty} captures the noise present in the observed sequences of training data. It arises when classes of the data present in the training data are overlapping and the hyper-plane separating these classes is not apparent as shown in Fig. \ref{Fig:uncertainty}.a. Increasing the amount of training data doesn't reduce the data uncertainty because the noise is embedded in the data itself.

\textbf{Epistemic Uncertainty or Model uncertainty} captures the confidence of the ML model in making predictions. It captures the state of limited knowledge that arises because of limited training data. Model uncertainty manifests itself in predictions made by the model on out-of-distribution data or regions in which the model is sparsely trained as shown in Fig. \ref{Fig:uncertainty}.b.

\subsection{Trustworthiness of ML algorithms}
A fundamental assumption of training a supervised ML algorithm is that the training data points and the new test data points are all sampled from the same distribution independently. However, the underlying probability distribution of the data may change over time, resulting in a mismatch between the distribution of the training data and the test data. Furthermore, the limited training data may not be sufficient to model the current data distribution \cite{najafabadi_deep_2015}. Such dataset shift scenarios severely degrade the accuracy and trustworthiness of the ML models when they are deployed. Uncertainty can be used to address this issue by providing insight into the limitations of the model's knowledge \cite{kendall2017uncertainties}. 

\subsection{Modelling uncertainty using Bayesian Principles}
Modeling uncertainty using Bayesian principles has been extensively studied before \cite{bayesian_uncertainty_original,kendall2017uncertainties,malinin2018predictive}. The uncertainty in the prediction of a classification model is given by the equation:
\begin{equation}
\label{eqn: intract_integral}
    P(y|\mathbf{x^{*}}, \mathcal{D}) = \int P(y|\mathbf{x^{*}}, \mathbf{\Theta} )P(\mathbf{\Theta}, \mathcal{D})d\mathbf{\Theta}
\end{equation}

where $y$ is the classification label, $\mathbf{x^{*}}$ is the test data, and $\mathcal{D}$ is the training dataset. A Bayesian model learns a distribution over the model parameters $\mathbf{\Theta}$ which is called the posterior over the model parameters represented by $P(\mathbf{\Theta}, \mathcal{D}) $. The predictive posterior ($P(y|\mathbf{x^{*}}, \mathcal{D})$) is generated using $P(\mathbf{\Theta}, \mathcal{D})$ given by Equation \ref{eqn: intract_integral}. 

\subsection{Ensembling techniques in Classical ML}
Ensemble is an ML technique that creates a classifier composed of a set of base-classifiers, that are then used for classification by taking a vote of their individual predictions \cite{10.1007/3-540-45014-9_1}. By creating an ensemble out of a diverse set of base-classifiers, the resulting algorithm reduces its reliance on one single classifier thereby reducing the risk of misclassification. In the domain of HMD, ensemble has been used for increasing the accuracy \cite{8280556,8465828}, and for providing defense against adversarial attacks \cite{kuruvila2020defending,8587644}.

\subsection{Related Work}
\label{sec:rel_work}
Ensemble methods have been employed in the ML community for increasing the accuracy of classification \cite{10.1007/3-540-45014-9_1}. It has been used to increase the accuracy of HMD \cite{8280556,8465828 }. It has been used for evading adversarial attacks on HMDs \cite{kuruvila2020defending, 8587644}. In the domain of deep learning, Lakshminarayanan et al. proposed an ensemble based approach to quantify the uncertainty in computer vision tasks  \cite{lakshminarayanan2017simple}. The base classifiers of the ensemble were obtained by random initialization of the Neural Network parameters. In practice, the deep learning models require a lot of training data, which is not available in the HMD community. Different from all previous work, our work evaluates the trustworthiness of the ML models  in HMDs using the proposed ensemble based uncertainty framework.

Chawla et al. performed a brief analysis on detection of unknown applications using DVFS signatures \cite{chawla_iot}. For evaluating the prediction probabilites, they have used Platt's scaling \cite{Platt99probabilisticoutputs}. They perform logistic regression on the output of the base classifier with respect to the true class labels to get an estimate of the prediction probabilities. However, the predictive probability generated at the output of the logistic activation function is often misconstrued as the model confidence. A model can make uncertain predictions with a high value of the logistic activation function \cite{gal2016dropout}. Passing a point estimate of the prediction through a logistic activation function may result in irrational high confidence predictions for data points lying in the region of overlap or the sparsely trained region. On the contrary, our method uses the entire distribution of the ensemble prediction to estimate the predictive uncertainty.  

Backes et al. used Bayesian ML to implement a malware detector for Android Operating System \cite{7961981}. They used the posterior over the parameters to explain the decisions of the ML model, and to increase the classification accuracy. While the same principles can be applied to HMD, the use of Bayesian ML will make the implementation harder and computationally slower, as compared to our proposed approach. Furthermore, we connect domain specific problems faced by HMD to well-known uncertainty problems.  

\section{Proposed Framework}
\label{sec:proposed_methodology}
\subsection{Mathematical Approach}
In this section, we propose an ensemble-based uncertainty estimator framework to evaluate the trustworthiness of the ML models employed in HMDs. The ensemble-based approach provides an estimate of the predictive uncertainty, with minor modifications to the conventional training pipeline.

Evaluating the integral in Equation \ref{eqn: intract_integral} is intractable. In general the integral is handled via sampling from $P(\mathbf{\Theta}, \mathcal{D}) $ leading to an ensemble of models formally denoted by:
\begin{equation}
\label{eqn:ensemble}
    \left \{ P(y|\mathbf{x^{*}}, \mathbf{\Theta^{(m)}}) \right \}_{m=1}^{M} , \mathbf{\Theta^{(m)}} \sim p(\mathbf{\Theta}|\mathcal{D})     
\end{equation}
With the sampled parameters, Equation \ref{eqn: intract_integral} is approximated as:
\begin{equation}
\label{eqn: ensemble_eqn}
    P(y|\mathbf{x^{*}}, \mathcal{D}) \approx  \frac{1}{M} \sum_{m=1}^{M}P(y|\mathbf{x^{*}}, \mathbf{\Theta^{(m)}}) , \mathbf{\Theta^{(m)}} \sim p(\mathbf{\Theta}|\mathcal{D}) 
\end{equation}

Equation \ref{eqn: ensemble_eqn} states that for creating a predictive distribution of the classification model, we average the outputs of the individual models created by sampling parameters from $P(\mathbf{\Theta}, \mathcal{D}) $. In practice, Bayesian ML approaches are computationally expensive and are difficult to implement as compared to the standard ML approaches. We propose to use ensembles for estimating the predictive uncertainty instead of the soft model selection denoted by $\mathbf{\Theta^{(m)}} \sim p(\mathbf{\Theta}|\mathcal{D})$. It should be noted that soft model selection using Bayesian principles differs from using ensembles that combine diverse models to create more powerful models \cite{Minka2002BayesianMA}. The predictive uncertainty is estimated by calculating the entropy of the predictive posterior given by the equation:
\vspace{-2mm}
\begin{equation}
\label{eqn:entropy}
H(Y) = -\sum_{i=1}^{n} P(y|\mathbf{x^{*}}, \mathcal{D})\log P(y|\mathbf{x^{*}}, \mathcal{D}) 
\end{equation}

\begin{figure*}[!ht]
   \begin{minipage}{0.59\textwidth}
     \centering
     \includegraphics[width=1\linewidth]{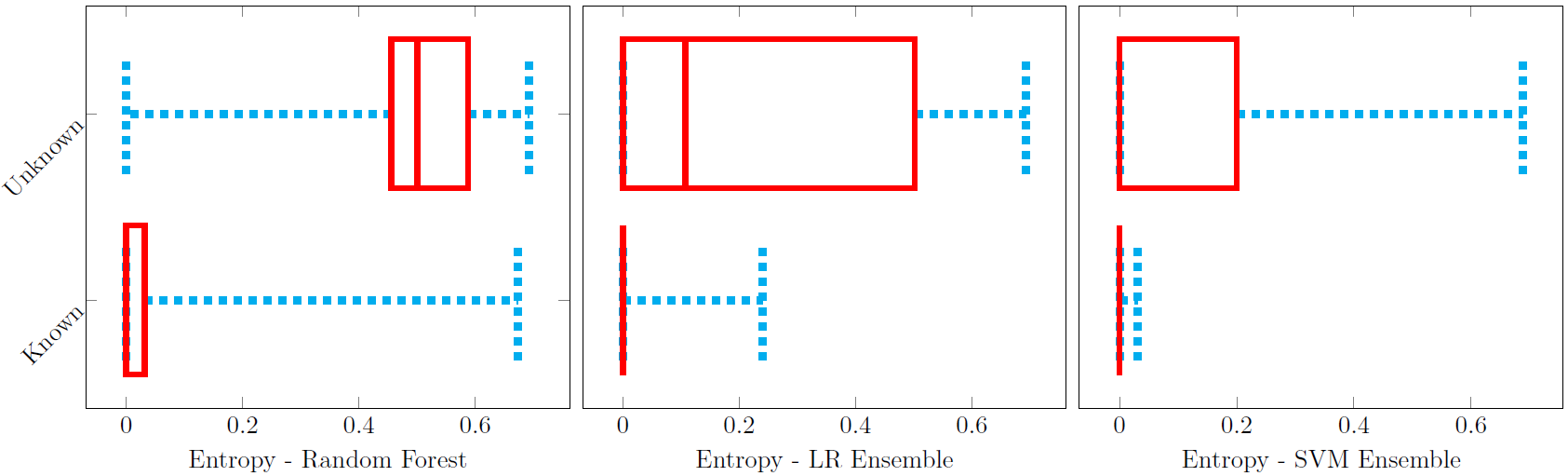}
     \caption{Estimated entropies for the DVFS Dataset}\label{Fig:box_dvfs}
   \end{minipage}\hfill
   \begin{minipage}{0.41\textwidth}
     \centering
     \includegraphics[width=1\linewidth]{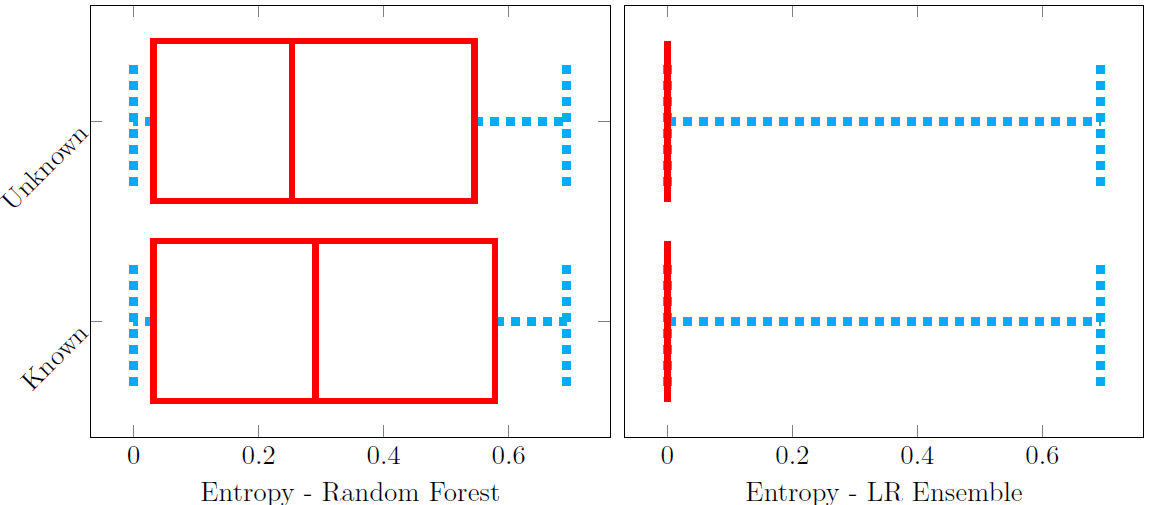}
     \caption{Estimated entropies for the HPC Dataset}\label{Fig:box_perf}
   \end{minipage}
\end{figure*}
\vspace{-2mm}
\subsection{Implementation details}
 Fig. \ref{Fig:overview} shows the proposed ensemble-based framework that is used to estimate the predictive uncertainty. In this work, we use the technique of bagging to create an ensemble of base-classifiers \cite{breiman_bagging_1996}. For every base-classifier, bagging generates a new training dataset by drawing samples from the original training dataset with replacement. The resulting training set is called a bootstrap replicate of the original training dataset. We propose to add an Uncertainty Estimator module, that creates a frequency distribution of the decisions of the base classifiers of the ensemble. The amount of dispersion in this distribution gives us an estimate of the predictive uncertainty of the model. Intuitively, we want to measure how much the predictions of base classifiers of the ensemble concur or differ from each other. Confident predictions on inputs that are close to the training dataset should have relatively less disagreement between the models in contrast to an out-of-distribution input that is far from the training data. We quantify this agreement/disagreement between the models by calculating the entropy of the frequency distribution of the decisions as formalized in Equation \ref{eqn:entropy}.

\subsection{Using Uncertainty to address trustworthiness issues impacting HMD}
\label{ssec: hypothesis}
Two major domain-specific problems affect the trustworthiness of the ML models in an HMD. First, its ability to handle zero-day malware. What is the output of the ML model when it encounters signatures from zero-day malware on which it has not been trained on? When the signature lies in a sparsely trained region, the model will make an erroneous prediction that undermines the trustworthiness of the deployed HMD. This problem can be addressed by obtaining the epistemic uncertainty of the ML model. If the trained model encounters out-of-distribution test data, \textit{i.e.}, signatures generated from completely new applications, or malware families, the predictive uncertainty will be high. Such a case is conveyed in Fig. \ref{Fig:uncertainty}.b where the test data lies in a sparsely trained region. The base classifiers of the ensemble tend to make random predictions on the unknown data as shown in Fig. \ref{Fig:overview}.a. The resulting randomness leads to high predictive uncertainty when compared to the predictions made on the in-distribution inputs, provided the decision boundaries are well defined. 

Second, the signatures used for training the ML model should form disjoint classes in the latent space of the training data. This is governed by factors like, the selection of hardware sensors used for collecting the data to train the ML models in HMD, the non-determinism observed in the signatures, and the background conditions while collecting the training data that is dictated by multiple workloads executing their instructions in the background. Test data points that lie in the region of overlap often produce incorrect predictions. This issue can be addressed by data uncertainty. As shown in Fig. \ref{Fig:uncertainty}.a, a high data uncertainty in the training data implies an absence of a well-defined decision boundary in the latent space of the data. In the case of Fig. \ref{Fig:uncertainty}.a, the three classification classes are overlapping. Consequently, the in-distribution test data shows high uncertainty. On the contrary, the predictive uncertainty for the in-distribution input is low in the case of well-defined decision boundaries. This gives us the trustworthiness of the model in terms of the confidence it shows for predicting in-distribution inputs and helps us to identify acceptable training data points that results in non-overlapping classes.

\begin{figure}[!b]
   \begin{minipage}{0.23\textwidth}
     \centering
     \includegraphics[width=1\linewidth]{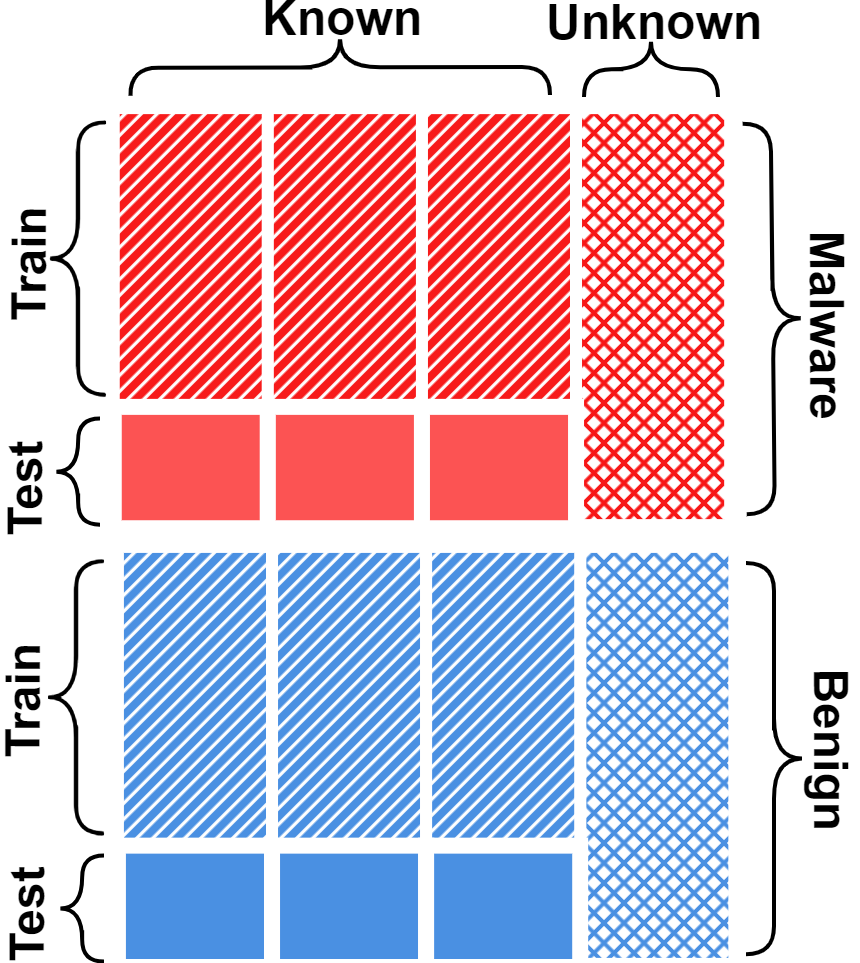}
     \caption{Dataset breakdown}\label{Fig:dataset}
   \end{minipage}\hfill
   \begin{minipage}{0.25\textwidth}
  \centering
  
        \resizebox{\textwidth}{!}{%
        \begin{tabular}{|c|c|}
        \hline
        \textbf{Split} & \textbf{\# of Samples} \\ \hline
        \multicolumn{2}{|c|}{\textbf{DVFS}} \\ \hline
        \textit{Train} & 2100 \\ \hline
        \textit{Test (Known)} & 700 \\ \hline
        \textit{Unknown} & 284 \\ \hline
        \multicolumn{2}{|c|}{\textbf{HPC}} \\ \hline
        \textit{Train} & 44605 \\ \hline
        \textit{Test (Known)} & 6372 \\ \hline
        \textit{Unknown} & 12727 \\ \hline
        \end{tabular}%
        }
    \captionof{table}{Dataset taxonomy}\label{tab:dataset}    
   \end{minipage}
\end{figure}

\section{Evaluation Methodology}
\label{sec:eval_method}
\subsection{Datasets under consideration}
We used two different datasets that are proposed in the literature. The first uses \textit{Dynamic Voltage and Frequency States (DVFS)}, which reflects power scaling behavior in modern SoCs \cite{chawla_iot,ncgt6_ncgt6securing-iot-devices-using-dynamic-power-management_2020}. Chawla et al. inferred applications running on Android platforms by extracting features from the time series of DVFS states. The authors were able to obtain an F1 score $> 0.88$ for classifying benign and malware workloads.

The second dataset is a HPC dataset \cite{hpc-myth,noauthor_bu-icsghardware_performance_counters_can_detect_malware_myth_or_fact_2020}. HPCs are dedicated registers built into a processor that keeps count of the number of microarchitectural events in a CPU core. Zhou et al. achieved an F1 score of up to $0.8$ for classification between benign and malware workloads for certain ML models.


\begin{figure*}[!htb]
  \begin{minipage}{0.5\textwidth}
     \centering
     \includegraphics[width=0.9\linewidth]{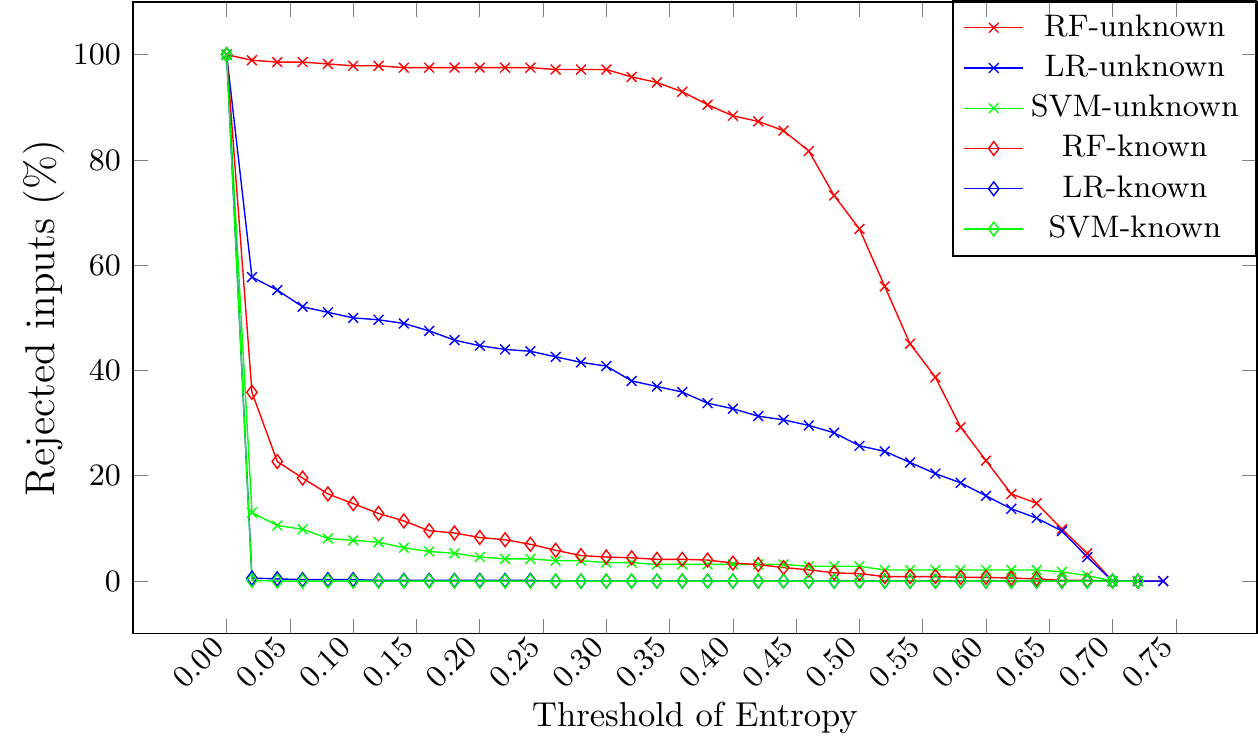}
  \end{minipage}
  \begin{minipage}{0.5\textwidth}
     \centering
     \includegraphics[width=0.9\linewidth]{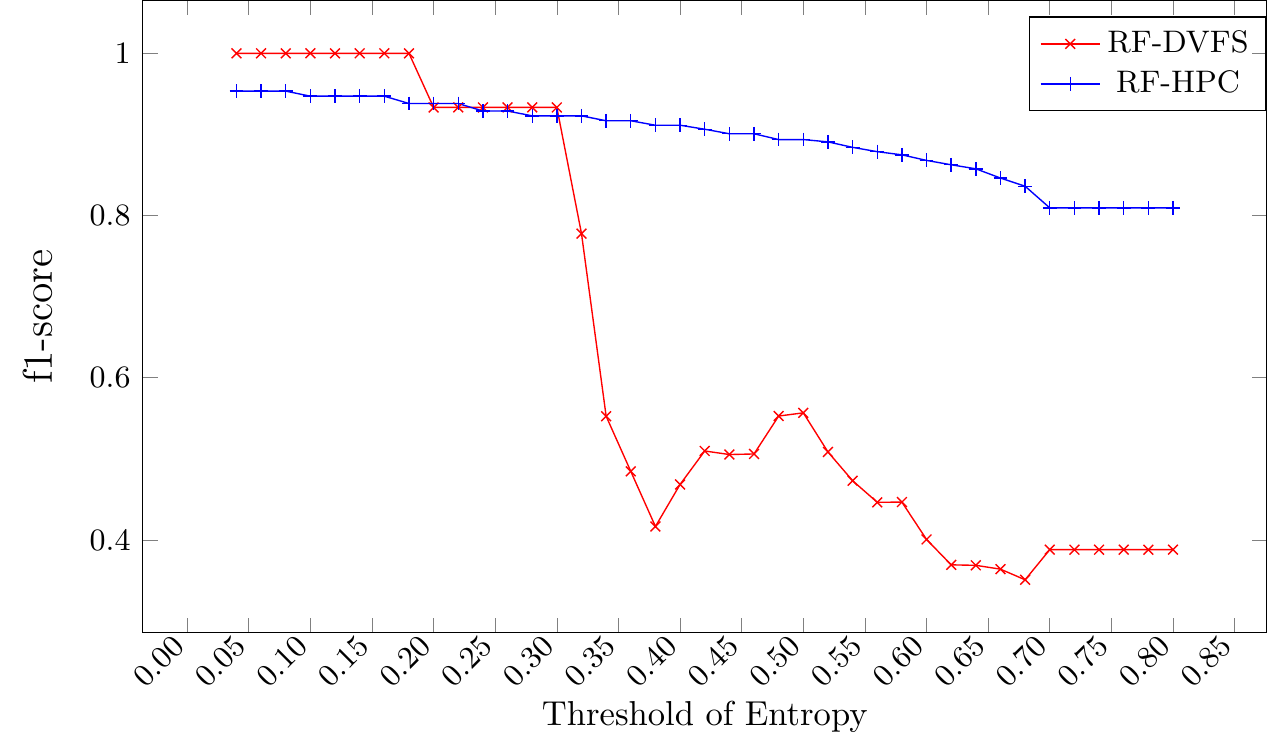}
  \end{minipage}
  \caption{(a) Rejected known/unknown data for DVFS dataset vs. threshold of entropy. (b) f1-score of the DVFS and HPC dataset vs. threshold of entropy}\label{fig:reject_input_dvfs}
\end{figure*}


\textbf{Known and Unknown data.} We first separate the signatures in the dataset on the basis of the applications they were derived from and placed these signatures into two buckets: known and unknown as shown in Fig. \ref{Fig:dataset}. The signatures from the known dataset were divided into training and testing datasets. The testing dataset is used to evaluate the uncertainties for the in-distribution data and the unknown dataset is used to evaluate the uncertainties of the out-of-distribution data. Table \ref{tab:dataset} shows the details of the individual datasets.

\subsection{Evaluation Framework}
The evaluation framework was developed on Python using the standard ML libraries provided by \verb|scikit-learn|. We used the ensemble framework provided by the sklearn libraries to generate the ensemble of base classifiers using the method of bagging. We used the \verb|estimators_| attribute of sklearn's ensemble framework to access the outputs of the base classifiers participating in the ensemble. The individual outputs were used for generating the frequency distribution used for calculating the entropy of the prediction. Generating the framework requires minor modifications to the training pipeline and can be performed using a small Python script.

\section{Evaluation results}
\label{sec:eval_results}

\subsection{DVFS dataset}

\begin{figure*}[!htb]
  \begin{minipage}{0.5\textwidth}
     \centering
     \includegraphics[width=0.9\linewidth]{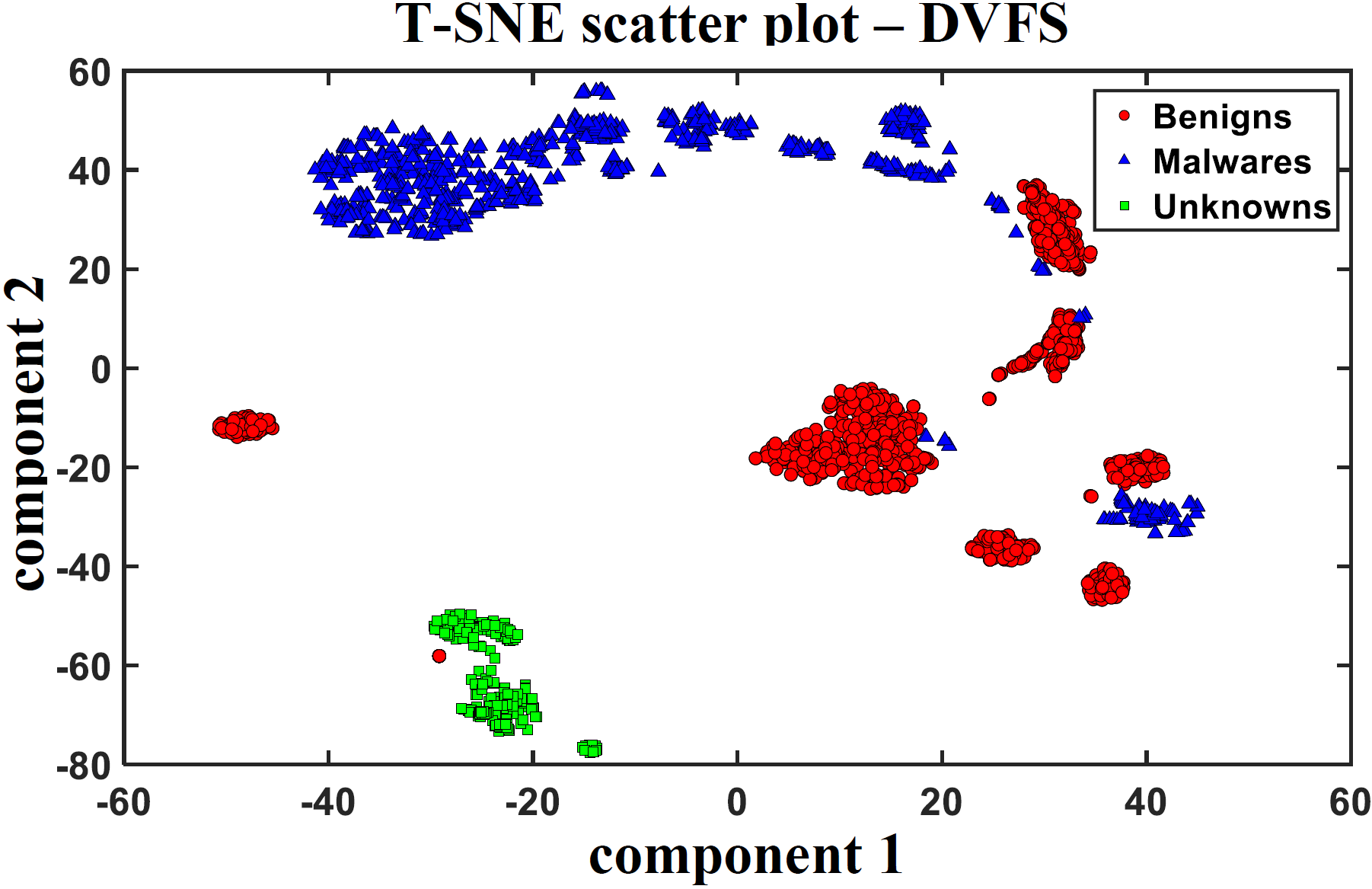}
  \end{minipage}
  \begin{minipage}{0.5\textwidth}
     \centering
     \includegraphics[width=0.9\linewidth]{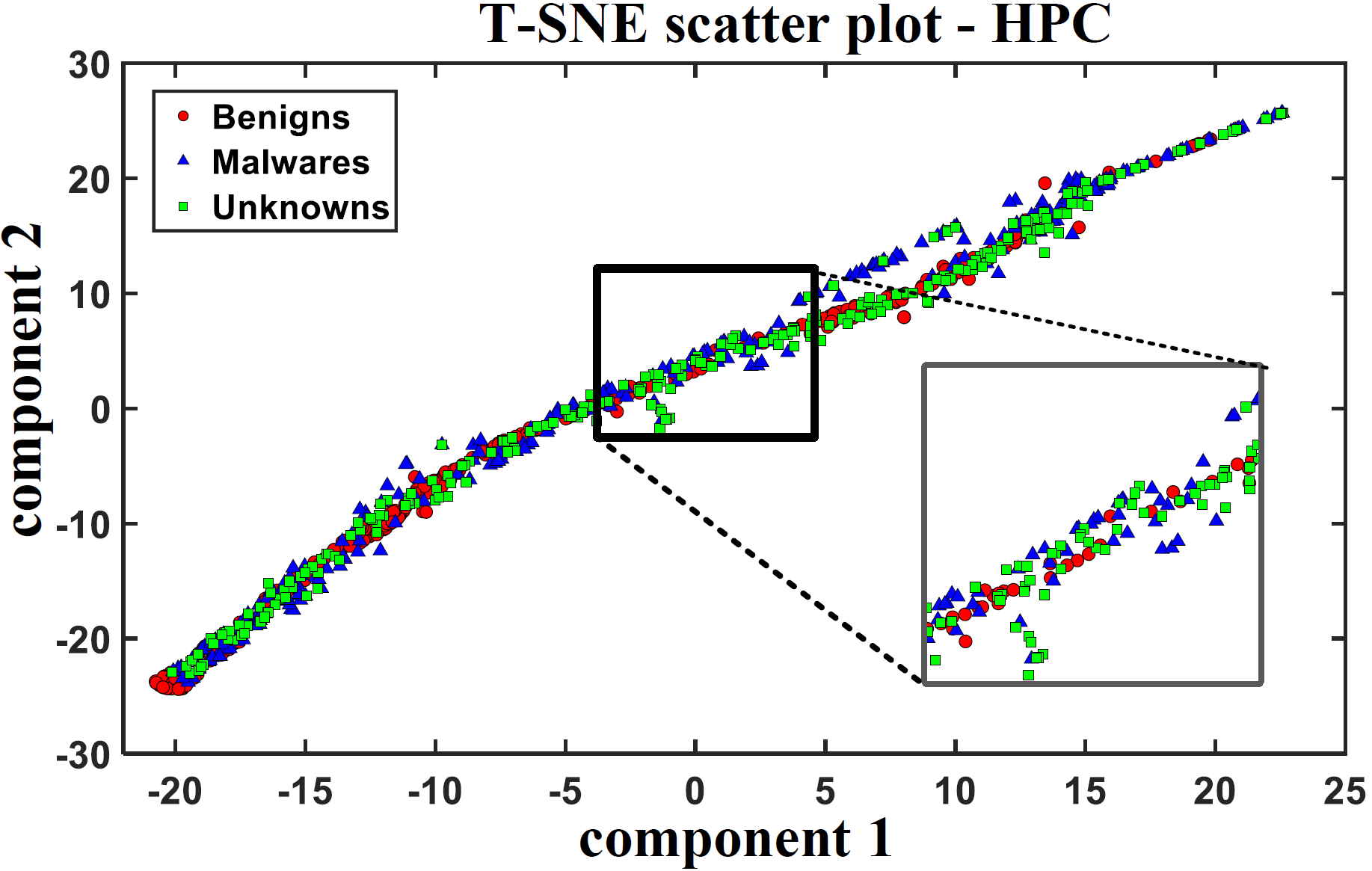}
  \end{minipage}
  \caption{t-SNE plot visualizing the Benign and Malware class in the training data, and the Unknown data (a) DVFS (b) HPC}\label{Fig:tsne}

\end{figure*}

Fig. \ref{Fig:box_dvfs} shows the boxplot of the distribution of estimated entropies for the predictions on the known and the unknown data. We obtain this distribution for the ensembles Random Forest, Logistic Regression, and SVM based ensemble. We can observe that predictions made on the unknown workloads have higher entropies than the known workloads. We can make two inferences from this observation: First, due to the low predictive uncertainties of the known data, we can say that the training data forms disjoint classes. Second, the high predictive uncertainty of the unknown data shows that the applications considered as part of the unknown dataset have signatures that are far away from training data.  We confirm this by plotting the t-SNE of the training data shown in Fig. \ref{Fig:tsne}.a to visualize the high dimensional latent space of the data.


\begin{figure}[!htb]
  \begin{minipage}{0.24\textwidth}
     \centering
     \includegraphics[width=1\linewidth]{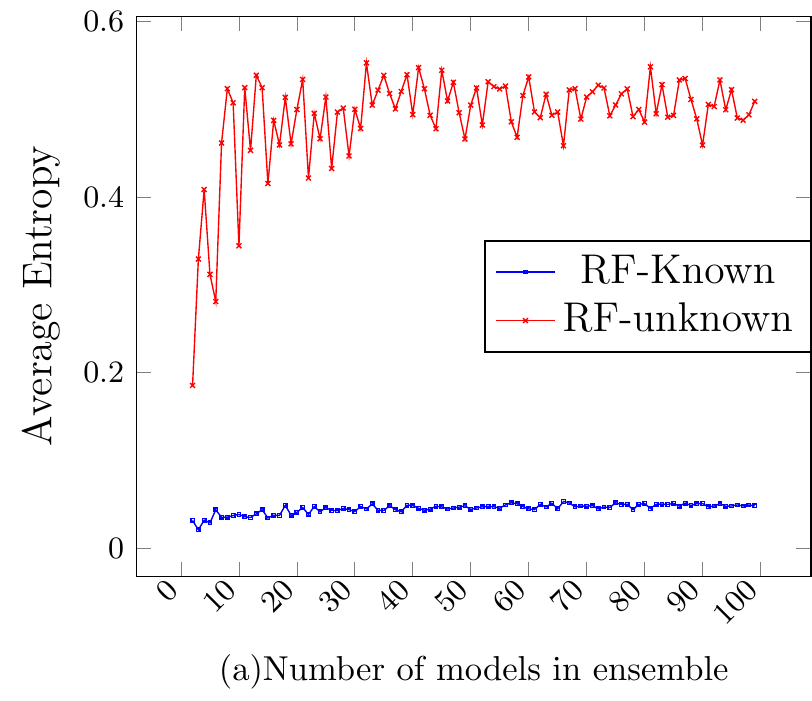}
  \end{minipage}
  \begin{minipage}{0.24\textwidth}
     \centering
     \includegraphics[width=1\linewidth]{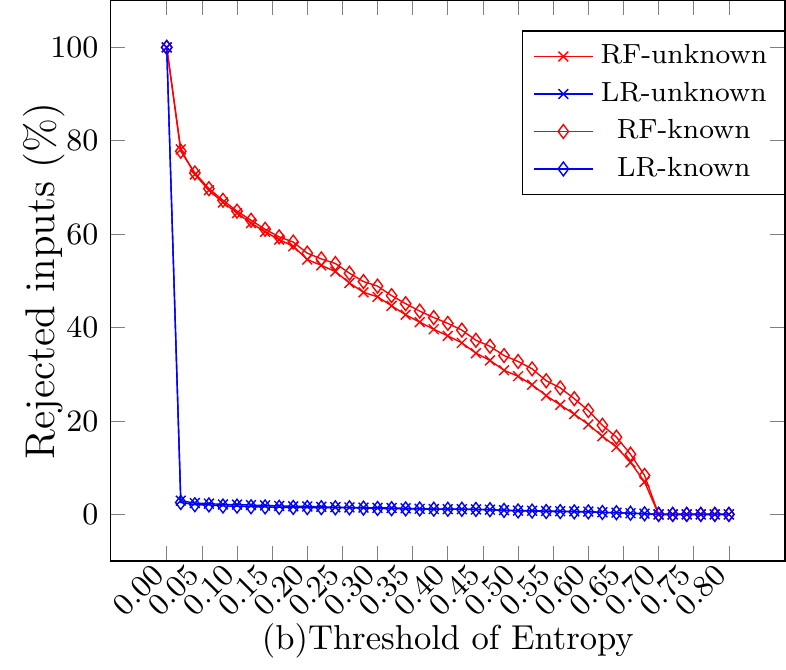}
  \end{minipage}
  \caption{(a) Average entropy vs \# of base-classifiers in ensemble. (b) Rejected known/unknown data for HPC dataset vs. threshold of entropy.}\label{Fig:merged_fig}

\end{figure}

 Fig. \ref{fig:reject_input_dvfs}.a shows the number of applications rejected by the classifier by varying the threshold of the entropy. If the entropy of a particular prediction goes beyond the threshold, we reject that decision citing the uncertainty in the prediction. Out of the three ensembles that we had considered, RF gives the best results, wherein, with a threshold of 0.40 we can reject about $95\%$ of the unknown workloads on which our model wasn't confident, at the same time rejecting $<5\%$ of the known workloads. The resulting increase in the f1-score is shown in Fig. \ref{fig:reject_input_dvfs}.b. SVM's performance was underwhelming in our experiments. With an uncertainty threshold of 0.04, we can reject only $40\%$ of unknown workloads with no rejections for known workloads. The ``quality" of uncertainty obtained in the case of SVM using our ensemble approach is quite poor. Since SVM is a convex optimization problem, bagging is unable to generate enough diversity in the base classifiers with the DVFS dataset, which explains the poor quality of uncertainty scores for the SVM ensemble.

Fig. \ref{Fig:merged_fig}.a shows the entropies estimated for the RF classifier as we increase the number of base-classifiers. Ensembles with base classifier as SVM and LR yield similar results, hence we show only RF for clarity. It can be observed that the estimated entropy stabilizes when the number of base-classifiers exceeds 20. Therefore, using more than 20 base-classifiers for measuring uncertainty adds unnecessary overhead for this dataset.

\subsection{Performance Counter (HPC) dataset}
SVM failed to converge using the bootstrapped dataset, therefore we do not include it in our observations. The boxplots in Fig. \ref{Fig:box_perf} shows that the estimated entropy for the known dataset is as high as the unknown dataset. This means that the ensemble is making highly uncertain predictions for the in-distribution data points. To explore this further, we plotted the t-SNE plot of the HPC training dataset. As can be observed in Fig. \ref{Fig:tsne}.b, the benign and malware classes in the HPC dataset are overlapping leading to high data uncertainty which is reflected in our estimate of the predictive uncertainty. Moreover, the unknown dataset has high predictive uncertainty because it lies in the region of class overlap and not the out-of-distribution region as we had hypothesized. We reject the unknown and the known workloads in the same fashion upon varying the threshold of the entropy as shown in Fig. \ref{Fig:merged_fig}.b. The ensemble based approach fails to identify whether the source of uncertainty is alleatoric or epistemic, which is one of its limitations. For the HPC dataset, even if we are able to get high accuracies on the known data ($84\%$ in case of RF as reported in the original paper), the model is not confident on these predictions. Upon rejecting the uncertain predictions, we can increase the f1-score of the RF model to 95\% as shown in Fig. \ref{fig:reject_input_dvfs}.b. The improvement is due to the increase in precision, \textit{i.e.}, the reduction in false positives. However, we also observe a reduction in recall.

\subsection{Summary of results}
The following can be concluded from our results:
\begin{itemize}
    \item We were able to successfully detect unknown applications in the DVFS dataset due to the out-of-distribution nature of the unknown data, verifying our hypothesis that model uncertainty can be used to detect zero-day malware.
    \item The high data uncertainty in the HPC dataset concurs with the conclusions of Zhou et al. \cite{hpc-myth}. Through our uncertainty analysis, we can conclude that the  HPC dataset presented in \cite{hpc-myth} cannot be used to train trustworthy ML model for HMD detection.
\end{itemize}

\vspace{-5mm}
\section{Conclusion}
\label{sec:conclusion}
In this paper, we propose an ensemble-based framework for estimating predictive uncertainties in HMDs. The predictive uncertainty is used to evaluate the trustworthiness of the HMDs. We demonstrate, using empirical results, both data uncertainty and model uncertainty encountered during the task of classifying a workload into benign or malware. We apply our framework to a DVFS-based HMD and an HPC-based HMD, proposed in the literature, to evaluate their robustness to zero day malwares and uncertain predictions. We discuss the limitations of our approach arising from the inability to separate the different sources of uncertainty. Tackling this issue will be the focus of our future work.  

\section{Acknowledgement}
\vspace{-1mm}
This material is based on work supported by Semiconductor Research Corporation and TxACE under \#2810.002
\vspace{-1mm}

\bibliographystyle{IEEEtran}
\bibliography{ref.bib}

\begin{thebibliography}{10}
\providecommand{\url}[1]{#1}
\csname url@samestyle\endcsname
\providecommand{\newblock}{\relax}
\providecommand{\bibinfo}[2]{#2}
\providecommand{\BIBentrySTDinterwordspacing}{\spaceskip=0pt\relax}
\providecommand{\BIBentryALTinterwordstretchfactor}{4}
\providecommand{\BIBentryALTinterwordspacing}{\spaceskip=\fontdimen2\font plus
\BIBentryALTinterwordstretchfactor\fontdimen3\font minus
  \fontdimen4\font\relax}
\providecommand{\BIBforeignlanguage}[2]{{%
\expandafter\ifx\csname l@#1\endcsname\relax
\typeout{** WARNING: IEEEtran.bst: No hyphenation pattern has been}%
\typeout{** loaded for the language `#1'. Using the pattern for}%
\typeout{** the default language instead.}%
\else
\language=\csname l@#1\endcsname
\fi
#2}}
\providecommand{\BIBdecl}{\relax}
\BIBdecl

\bibitem{demme_performance_counter}
J.~Demme, M.~Maycock, J.~Schmitz, A.~Tang, A.~Waksman, S.~Sethumadhavan, and
  S.~Stolfo, ``On the feasibility of online malware detection with performance
  counters,'' ser. ISCA '13.\hskip 1em plus 0.5em minus 0.4em\relax ACM, 2013.

\bibitem{8280556}
K.~N. {Khasawneh}, M.~{Ozsoy}, C.~{Donovick}, N.~{Abu-Ghazaleh}, and
  D.~{Ponomarev}, ``Ensemblehmd: Accurate hardware malware detectors with
  specialized ensemble classifiers,'' \emph{IEEE Transactions on Dependable and
  Secure Computing}, vol.~17, no.~3, pp. 620--633, 2020.

\bibitem{8465828}
H.~{Sayadi}, N.~{Patel}, S.~M. {P.D.}, A.~{Sasan}, S.~{Rafatirad}, and
  H.~{Homayoun}, ``Ensemble learning for effective run-time hardware-based
  malware detection: A comprehensive analysis and classification,'' 2018, pp.
  1--6.

\bibitem{8192483}
A.~{Nazari}, N.~{Sehatbakhsh}, M.~{Alam}, A.~{Zajic}, and M.~{Prvulovic},
  ``Eddie: Em-based detection of deviations in program execution,'' in
  \emph{2017 ACM/IEEE 44th Annual International Symposium on Computer
  Architecture (ISCA)}, 2017, pp. 333--346.

\bibitem{chawla_iot}
N.~{Chawla}, A.~{Singh}, H.~{Kumar}, M.~{Kar}, and S.~{Mukhopadhyay},
  ``Securing iot devices using dynamic power management: Machine learning
  approach,'' \emph{IEEE Internet of Things Journal}, pp. 1--1, 2020.

\bibitem{noauthor_qualcomm_2017}
\BIBentryALTinterwordspacing
``\BIBforeignlanguage{en}{Qualcomm {Mobile} {Security}},'' Jan. 2017. [Online].
  Available: \url{https://www.qualcomm.com/products/features/mobile-security}
\BIBentrySTDinterwordspacing

\bibitem{kendall2017uncertainties}
A.~Kendall and Y.~Gal, ``What uncertainties do we need in bayesian deep
  learning for computer vision?'' 2017.

\bibitem{lakshminarayanan2017simple}
B.~Lakshminarayanan, A.~Pritzel, and C.~Blundell, ``Simple and scalable
  predictive uncertainty estimation using deep ensembles,'' 2017.

\bibitem{gal2016dropout}
Y.~Gal and Z.~Ghahramani, ``Dropout as a bayesian approximation: Representing
  model uncertainty in deep learning,'' 2016.

\bibitem{malinin2018predictive}
A.~Malinin and M.~Gales, ``Predictive uncertainty estimation via prior
  networks,'' 2018.

\bibitem{najafabadi_deep_2015}
M.~M. Najafabadi, F.~Villanustre, T.~M. Khoshgoftaar, N.~Seliya, R.~Wald, and
  E.~Muharemagic, ``Deep learning applications and challenges in big data
  analytics,'' \emph{Journal of Big Data}, vol.~2, p.~1, Feb.

\bibitem{bayesian_uncertainty_original}
J.~L. Beck and L.~S. Katafygiotis, ``Updating models and their uncertainties.
  i: Bayesian statistical framework,'' \emph{Journal of Engineering Mechanics},
  vol. 124, no.~4, pp. 455--461, 1998.

\bibitem{10.1007/3-540-45014-9_1}
T.~G. Dietterich, ``Ensemble methods in machine learning,'' in \emph{Multiple
  Classifier Systems}.\hskip 1em plus 0.5em minus 0.4em\relax Berlin,
  Heidelberg: Springer Berlin Heidelberg, 2000, pp. 1--15.

\bibitem{kuruvila2020defending}
A.~P. Kuruvila, S.~Kundu, and K.~Basu, ``Defending hardware-based malware
  detectors against adversarial attacks,'' 2020.

\bibitem{8587644}
K.~N. {Khasawneh}, N.~B. {Abu-Ghazaleh}, D.~{Ponomarev}, and L.~{Yu},
  ``Adversarial evasion-resilient hardware malware detectors,'' in \emph{2018
  IEEE/ACM International Conference on Computer-Aided Design (ICCAD)}, 2018,
  pp. 1--6.

\bibitem{Platt99probabilisticoutputs}
J.~C. Platt, ``Probabilistic outputs for support vector machines and
  comparisons to regularized likelihood methods,'' in \emph{ADVANCES IN LARGE
  MARGIN CLASSIFIERS}.\hskip 1em plus 0.5em minus 0.4em\relax MIT Press, 1999,
  pp. 61--74.

\bibitem{7961981}
M.~{Backes} and M.~{Nauman}, ``Luna: Quantifying and leveraging uncertainty in
  android malware analysis through bayesian machine learning,'' in \emph{2017
  IEEE European Symposium on Security and Privacy (EuroS P)}, 2017, pp.
  204--217.

\bibitem{Minka2002BayesianMA}
T.~Minka, ``Bayesian model averaging is not model combination,'' 2002.

\bibitem{breiman_bagging_1996}
\BIBentryALTinterwordspacing
L.~Breiman, ``Bagging {Predictors},'' \emph{Machine Learning}, vol.~24, no.~2,
  pp. 123--140, Aug. 1996. [Online]. Available:
  \url{https://doi.org/10.1023/A:1018054314350}
\BIBentrySTDinterwordspacing

\bibitem{ncgt6_ncgt6securing-iot-devices-using-dynamic-power-management_2020}
\BIBentryALTinterwordspacing
ncgt6,
  ``ncgt6/{Securing}-{IoT}-{Devices}-using-{Dynamic}-{Power}-{Management},''
  Nov. 2020, original-date: 2020-11-21T01:16:28Z. [Online]. Available:
  \url{https://github.com/ncgt6/Securing-IoT-Devices-using-Dynamic-Power-Management}
\BIBentrySTDinterwordspacing

\bibitem{hpc-myth}
B.~Zhou, A.~Gupta, R.~Jahanshahi, M.~Egele, and A.~Joshi, ``Hardware
  performance counters can detect malware: Myth or fact?'' ser. ASIACCS
  '18.\hskip 1em plus 0.5em minus 0.4em\relax New York, NY, USA: ACM, 2018.

\bibitem{noauthor_bu-icsghardware_performance_counters_can_detect_malware_myth_or_fact_2020}
\BIBentryALTinterwordspacing
Aug. 2020, original-date: 2018-03-15T22:16:06Z. [Online]. Available:
  \url{https://github.com/bu-icsg/Hardware\_Performance\_Counters\_Can\_Detect\_Malware\_Myth\_or\_Fact}
\BIBentrySTDinterwordspacing

\end{thebibliography}

\end{document}